\documentclass[12pt]{article}

\usepackage{latexsym}
\usepackage{amssymb}
\usepackage{amsfonts}

\usepackage{hyperref}

\def\nc#1{\newcommand{#1}}
\def\rnc#1{\renewcommand{#1}}

\def\a{\alpha}
\def\b{\beta}
\nc{\g}{\gamma}
\def\d{\delta}
\nc{\D}{\Delta} 
\nc{\e}{\eta}
\nc{\ep}{\epsilon}

\nc{\ve}{\varepsilon}
\nc{\G}{\Gamma}

\nc{\la}{\lambda}
\nc{\La}{\Lambda}
\nc{\om}{\omega}
\nc{\Om}{\Omega}
\nc{\vphi}{\varphi}
\nc{\si}{\sigma}
\nc{\Si}{\Sigma}
\rnc\th{\theta}
\nc\Th{\Theta}
\nc{\z}{\zeta}

\def\cL{{\cal L}}

\def\cN{{\cal N}}

\nc{\got}[1]{\mathfrak{#1}} 

\def\det{{\rm det}}

\nc\im{{\rm Im}\, }
\nc\re{{\rm Re}\, }
\def\tr{{\rm tr}}
\def\Tr{{\rm Tr}}

\nc{\Rt}{{\tilde R}}
\nc{\CC}{{\mathbb C}}
\nc{\RR}{{\mathbb R}}
\nc{\HH}{{\mathbb H}}
\nc{\NN}{{\mathbb N}}
\nc{\ZZ}{{\mathbb Z}}
\nc{\MM}{{\mathbb M}}

\nc{\eql}{\eqalign}
\nc{\dis}{\displaylines}
\nc{\ce}{\centerline}
\nc{\hf}{\hspace{\fill}}
\nc{\hs}{\hspace*}
\nc{\vs}{\vskip .3cm}
\nc{\non}{\nonumber\\}
\def\nn{\nonumber}

\nc{\noi}{\noindent}

\nc{\p}{\partial}
\nc{\na}{\nabla}
\def\x{\times}

\def\w{\wedge}

\nc{\beq}{\begin{equation}}
\nc{\eeq}{\end{equation}}
\nc{\beqa}{\begin{eqnarray}}
\nc{\eeqa}{\end{eqnarray}}
\nc{\beqas}{\begin{eqnarray*}}
\nc{\eeqas}{\end{eqnarray*}}
\nc{\barr}{\begin{array}}
\nc{\earr}{\end{array}}
\nc{\ben}{\begin{enumerate}}
\nc{\een}{\end{enumerate}}
\nc{\bit}{\begin{itemize}}
\nc{\eit}{\end{itemize}}

\nc{\sfrac}[2]{{\mbox{\large $\frac{#1}{#2}\,$}}}
\nc\half{\sfrac{1}{2}}
\nc\nfrac[2]{\mbox{\small $\frac{#1}{#2}$}}

\nc{\eq}[1]{\stackrel{\ref{#1}}{=}}

\nc{\oto}{\leftrightarrow}

\def\refeq#1{{(\ref{#1})}}

\nc{\twovec}[2]{{\scriptscriptstyle \left( 
\begin{array}{c} #1\\[-.2cm]  #2 \end{array}\right)}}
\nc{\twomat}[4]{\left( \begin{array}{cc} #1&#2\\ 
#3&#4\end{array}\right)}

\makeatletter
\@addtoreset{equation}{section}

\def\sec{\section}
\def\ssec{\subsection}

\nc{\ma}{m_A}
\nc{\lapm}{\la^{(\pm)}}
\nc{\fpm}{{f^{(\pm)}}}\nc{\fmp}{{f^{(\mp)}}}
\nc{\tQ}{{\tilde Q}}
\nc{\ov}[1]{\overline{#1}}
\nc\lat{{\tilde \la}}
\nc\Lat{{\tilde \La}}
\nc\ct{{\tilde c\,}}

\nc\Db{{\overline D3}}

\author{Jacek Pawe{\l}czyk\\[7mm]
\normalsize{\textit{Institute of Theoretical Physics}},
\normalsize{\textit{Warsaw University}}\\[-2mm]
\normalsize{\textit{ Ho\.za 69, PL-00-681 Warsaw, Poland}}\\
{\normalsize {\tt jacek.pawelczyk@fuw.edu.pl}}}

\begin{document}


\title{\bf \LARGE Gauge Symmetry Breaking\\
in a Throat Geometry\footnote{Work supported by  Polish Ministry of Science MNiSW 
under contract N202 176 31/3844 (2006-2008).}}
\maketitle \vskip 1cm

\begin{abstract} \normalsize
\noindent We analyze behaviour of D3-branes in BGMPZ throat geometry. We show that although single brane has some of the moduli stabilized multi-brane system tends to expand and form a
bound state. Such a system loses non-abelian gauge symmetry.

\end{abstract}
\thispagestyle{empty}

\bigskip
\newpage
\setcounter{page}1
\tableofcontents

\newpage

\sec{Introduction}

Flux compactifications \cite{GKP} provide numerous examples of modern string unification in which D-branes are commonly used  to provide gauge degrees of freedom. It is always necessary to stabilize branes moduli in order to get rid of massless scalars.
Stabilization mechanism must lead to realistic gauge symmetry e.g. color SU(3) should be unbroken. 

It may happen that the stabilization mechanism works properly for single brane configuration but fails for multi-brane systems.
The reason is that such systems can expand to form  non-trivial bound states \cite{myers,my}.  If so the gauge symmetry of  the multi-brane system is lost. 
This might be disastrous in models where D3-branes
provide color SU(3) degrees of freedom (see e.g. \cite{antoniadis}) but could appear blissful for
electroweak symmetry breaking.


In this paper we shall discuss
an  example realizing the above mentioned scenario. We shall consider D3-branes moving in 6d 
background with fluxes
of IIB superstring. The world-volume of the branes will
 totally cover 4d Minkowski space.
The possibility opens up due to Myers effect \cite{myers}.
The effect is suppressed for ISD $G_3$ fluxes e.g. 
 position moduli of a D3 brane on such a compact 6d manifold are massless
and moreover several D3-branes  do not interact with each other.
  This means that in order to have interacting D3-brane system
  we need to depart from ISD condition. 
  Here we shall concentrate on the BGMPZ throat 
\cite{BGMPZ} as an example of non-ISD $G_3$ background. Single D3 in this background breaks SUSY and tends to rest at the bottom on the throat where its 3 out of 6 moduli are
stabilized \cite{DKS}. 
We shall show that several D3-branes behave differently. 
They may expand and form fuzzy sphere $S^2$ \cite{johnm} or fuzzy $S^2\x S^2$ due to Myers effect. In fact we shall discuss slightly more general background
characterized by one free parameter and possibly having all moduli massive. Thus the obtained results can be also applied to different geometries.

The paper is organized as follows. In the next section we shall discuss string background we are going to work with. The content of the section is based on \cite{BGMPZ}.
 In Sec.\ref{myers-a} we shall derive the matrix dynamics of multi-D3-brane
  system based on \cite{myers} and analyze its behaviour. 
Sec.\ref{sec:sol}  presents solutions of the matrix equations of motion and discusses their properties and 
Sec.\ref{sec:sol-e} presents generalization of the result for the case of all massive moduli. Conclusions contains  remarks concerning unification model building aspects of the mechanism discussed. 
In Appendix we discuss some details  of the BGMPZ background and present description of approximations used.


\sec{The background geometry}

We shall work with BGMPZ background \cite{BGMPZ}. The latter is a deformation of KS background \cite{KS} and it will be treated here in the approximation of a small deviation from the latter. The parameter of deviation will be denoted by $q$.  
In this paper we shall concentrate on dynamics close to the bottom of the throat
i.e. $\tau\approx 0$ where $\tau$ is the radial coordinate of 
the 6d throat geometry. Here we shall describe approximations we are going to make. For more detailed description of the BGMPZ background we refer to the original work \cite{BGMPZ} and the Appendix. 

The D3-brane dynamics is non-trivial due to deviation from ISD of $G_3$ of the KS background. Since we are interested in the leading terms in the deviation parameter $q$ it is enough to take 
10d metric in KS form.  Close to the bottom of the throat ($\tau\approx 0$)
the 10d metric can be approximated as follows
\beq
ds^2_{10}=e^{2A_0}dx_4^2+\La^2 ds^2_6
\eeq
where $\La^2=g_s M\a'6^{-1/3}a_0^{1/2},\ e^{2A_0}=\ep^{4/3}(g_s M\a'2^{1/3}a_0^{1/2})^{-1}$.
The 6d manifold with the metric
\beq\label{6d}
ds^2_6=d\tau^2+\sfrac{\tau^2}2(g_1^2+g_2^2)+d\Om_3^2
\eeq
is topologically  a certain bundle  $\RR_+\ltimes S^2\ltimes S^3$, where $\RR_+$ is radial direction parameterized by $\tau$.
We focus on a such a small region of $S^3$ which can be approximately treated to be flat i.e.
$d\Om_3^2\approx \sum d\Th_i^2$ \footnote{For some details of the approximation see App.\ref{approx}}. In this approximation
$(g_1^2+g_2^2)$ is the metric on $S^2$ of the radius $\sqrt{2}$ thus $d\tau^2+\sfrac{\tau^2}2(g_1^2+g_2^2)$ becomes flat metric 
$\sum dX_i^2$ on $\RR^3$. Altogether
\beq\label{flat-6d}
ds^2_6\approx \sum dX_i^2+ d\Th_i^2
\eeq
The expression for the dilaton is given in App.\ref{sec:flatS}. In our conventions  (following \cite{DKS} above Eq.(12.3)) the string coupling is
$
g_s e^{\phi(\tau)}
$
with $\phi\to 0 $ for $\tau\to\infty$.

We also need  expressions for the fluxes. One easily derives (see e.g. \cite{DKS} Eq. (12.5))
\beq
C_4={g_s\!\!}^{-1}\,dV^4={g_s\!\!}^{-1}\,e^{4A}d^4x
\eeq
For the multi-D3-brane system we also need the expression for 
$C_6+B\w C_4$. $C_6$ is  defined through
\beqa
\tilde F_7=-*\tilde F_3= -dV^4\w *_6 F_3
\eeqa
with $\tilde F_n\equiv dC_{n-1}+H\w C_{n-3}$. From this we get
\beq
d(C_6+B\w C_4)=-dV^4\w(*_6F_3+{g_s\!\!}^{-1}\,H)+{g_s\!\!}^{-1}\,d(dV^4\w B)
\eeq
The first term on the r.h.s. must be an exact form too thus 
\beq
e^{4A}(*_6F_3+{g_s\!\!}^{-1}\,H)=d\om_{FH}
\eeq
for a 2-form $\om_{FH}$. Integrating this equation at the leading order at $\tau$ and $q$ we get
\beqa
\om_{FH}&=&-3e^{4A_0}Pq^2 \lat^2\ \tau(e_1+e_3)\w(e_2+e_4)
\eeqa
In the flat approximation (i.e. when 
\refeq{flat-6d} holds) we have (see App.\ref{sec:flatS})
\beqa
\om_{FH}=-\sfrac32 e^{4A_0}Pq^2 \lat^2\ \ep_{ijk}X_i\ d\Th_j\w d\Th_k
\eeqa
Thus  one gets
\beq
C_6+B\w C_4=-d^4x \w \om_{FH}+{g_s\!\!}^{-1}\,dV^4\w B
\eeq
with the above $\om_{FH}$.

\sec{Multi-D3 dynamics}
\label{myers-a}
The basis for the multi-brane dynamics is the 
Myers action
\cite{myers}. Her we present it in the form suitable for D3-branes which world-volume is totally aligned with 4d-Minkowski space-time.
The basic matrix-valued variables $\Phi_i$'s are non-abelian avatars of coordinates on 6d manifold orthogonal to the branes world-volume.
\beqa\label{myers}
S=-T_3\int\Tr\bigg[ e^{-\phi} dV^4\ \sqrt{\det(Q)}- { g_s} P[C_4+
i\la i_\Phi i_\Phi(C_6+C_4\w B)
\bigg]\nn
\eeqa
where $P$ means pull-back to the brane world-volume,
 $\la=2\pi\a'$ and 
\beqa
Q^i{}_j&=&\d^i{}_j+i\la[\Phi^i,\Phi^k](G_{kj}+B_{kj})\non
\label{detQ-e}
\sqrt{\det(Q)}&=&1+\sfrac i2\la\, [\Phi^i,\Phi^k]B_{ki}+
\sfrac14 \la^2\, [\Phi^i,\Phi^k][\Phi^j,\Phi^l]G_{kj}G_{li}+...
\eeqa
where $...$ means higher order terms in $[\Phi^i,\Phi^k]$ and $O(B^2)$.

In our conventions $\Phi_i$'s  are dimensionless thus $\a'$ is the  only dimensionful quantity and it  must drop from the action. 
Keeping only the terms displayed in formulae \refeq{myers} and \refeq{detQ-e} we get
\beqa\label{myers-my}
S&\approx & 
-
-T_3\int d^4x\,\Tr\Big[(e^{-\phi}-1)\ e^{4A}+
i\la\, i_\Phi i_\Phi ((e^{-\phi}-1)\ e^{4A} B+\,g_s\,\om_{FH})\non
&&\hspace{5cm} - \sfrac{\la^2}4 e^{-\phi}\ e^{4A}  [\Phi^i,\Phi^j][\Phi^k,\Phi^l]G_{ik}G_{kl}\ \Big]
\eeqa
where $G_{ij}=\La^2 \d_{ij}$

\ssec{Branes in the approximate geometry}


In the flat limit (App.\ref{approx}) (recall $\Phi_i\in\{X_i,\Th_i\})$
\beq\label{lag-eff-0}
\cL= -T_3 e^{4A_0}\tr\left(e_0+m^2 X_i^2 
+i\ct\, \ep_{ijk} X_i\, \Th_j\,\Th_k
-\frac{\Lat^2}4 [\Phi_i,\Phi_j]^2
 \right)
\eeq
where $\ct= \frac98 g_s M q^2 \lat^2,\ e_0=\sfrac{9q^2 \lat^2}{2},\ m^2=\frac94 q^4\lat^2,\ \Lat^2=g_s M\,6^{-1/3}a_0^{1/2}$.
After some rescalings of variables $\Phi$ the relevant terms of the  Lagrangian are
\beq\label{lag-eff}
\cL= -T_3 e^{4A_0}\tr\left(e_0+ \sfrac{m^4}{\La^4}\D e \right)
\eeq
with
\beqa\label{lag-eff0}
\D e&=&X^2 
+ic_1\, \ep_{ijk} X_i\, \Th_j\,\Th_k-\sfrac14 [X_i,X_j]^2-\sfrac12 [X_i,\Th_j]^2-\sfrac14 [\Th_i,\Th_j]^2\\
\label{c1}
&&c_1=\frac{\ct}{m^2\Lat^2}
\eeqa
It appears that for BGMPZ geometry $c_1=\frac32$. Notice that the dynamical part
of \refeq{lag-eff0} 
 depends only on $c_1$.
One can imagine  that  $m,\ \ct,\ \Lat$ takes different values in string geometries other then 
BGMPZ geometry so hereafter we shall keep $c_1$ as a free parameter keeping in mind more general applications.

 The e.o.m. one gets are
\beqa
&&2 X_i-[X_j,[X_i,X_j]]-[\Th_j,[X_i,\Th_j]]+i\sfrac{c_1}2 \ep_{ijk}[\Th_j,\Th_k]=0\non[-.5cm]
&&\label{eqm}\\[-.25cm]
&&-[X_j,[\Th_i,X_j]]-[\Th_j,[\Th_i,\Th_j]]+ic_1\ep_{ijk}[X_j,\Th_k]=0
\nn
\eeqa
In the following we are going to analyze solutions of its equations of motion.
Let us make some general remarks first. The Lagrangian \refeq{lag-eff0} is $SO(3)$ symmetric with  $X_i,\ \Th_i$ transforming as vectors. 
Notice that $X_i$'s are massive while  $\Th$'s are massless in agreement with results of \cite{DKS}. Then
the Lagrangian \refeq{lag-eff0} and e.o.m. \refeq{eqm} are invariant under a constant shift $\Th_i\to\Th_i+v_i$. Thus solutions go in families which differ by the vectors $v_i$. We shall  suppress these vector from farther considerations.
It it also easy to see that
$\ZZ_2:(X_i,\Th_i)\to -(X_i,\Th_i)$ changes the sign of $c_1$ in \refeq{lag-eff0} and in \refeq{eqm}. It means that the set of
solutions is $\ZZ_2$ invariant i.e. if there exist
one with $c_1>0$ there is another one with $c_1\to -c_1$ and
$(X_i,\Th_i)\to -(X_i,\Th_i)$. We are going to use silently this fact in the following sections. Notice also that the non-trivial repulsive force comeing from c$_1\, \ep_{ijk} X_i\, \Th_j\,\Th_k$ couple $X$'s and $\Th$'s thus non-trivial solutions must depends non-trivially at least on one $X$. The latter is massive so we expect there is always a critical value of $|c_1|$ below which there in no solution.

\ssec{Solutions}
\label{sec:sol}

First we are looking for solutions preserving the $SO(3)$ symmetry of 
 \refeq{lag-eff} thus having the general form:
\beqa\label{ans}
&&\hskip-1cm{}[X_i,X_j]=i \ep_{ijk}(a\ X_k+b\ \Th_k)\ ,\quad
{}[\Th_i,\Th_j]=i \ep_{ijk}(c\ X_k+d\ \Th_k)\ ,\quad
{}[X_i,\Th_j]=i \ep_{ijk}(f\ X_k+g\ \Th_k)\non
&&
\eeqa
Because the commutators must respect Jacobi identities it is much more convenient
 to change variables:
$X_i=\a\ l_i+\b\ r_i,\ \Th_i=\g\ l_i +\d\ r_i$, where $[l_i,r_j]=0,\ \cN=\tr(l_i l_i)=\tr(r_i r_i)$ (no sum over $i$) and
$l_i\ ( r_i)$ respect standard SU(2) commutation relations.
With the ansatz \refeq{ans} the e.o.m. \refeq{eqm} splits into 
independent  equations on $(\a,\g)$: $\a^2\g+\g^3-c_1 \a\g =0,\
\a^3+\a+\a\g^2-\half \g^2=0$. The equations are invariant under
$\ZZ_2:\ \g\to-\g$ thus the set of solutions also has this symmetry.
The same equations hold for $(\b,\d)$.
Then modulo the above $\ZZ_2$ factors, we get two classes of  solutions:
\beqa
(a)\quad&&\a=y, \quad\g=x, \quad\b=0,  \quad\d= 0\\
(b)\quad&&\a=y, \quad\b=y, \quad\g= x, \quad\d= x,
\eeqa
where
\beq
x=\frac{{\sqrt{2\left( \,{{c_1}}^2 + 1 \right) \,
       \left( {{c_1}}^2 - 2 \right) }}}{3\,{c_1}}\ ,\quad
y=\frac{{{c_1}}^2 - 2}{3\,{c_1}}
\eeq
One must note that there is a critical value of $c_1=\sqrt{2}$ below which there is no non-trivial solution.
The solutions (a) form representations of SU(2) and thus they are fuzzy sphere $S^2$ embedded in the bottom of the throat.
In the case (b) we obtain
SU(2)$\x$ SU(2) representations. This means that geometrically the latter can be thought as fuzzy $S^2\x S^2$.

The above configurations  have lower energy densities compare to $e_0$\footnote{We give only the value of the part of the energy \refeq{lag-eff} under the trace.} by
\beq\label{en1}
(a)\quad \D e=-\cN\, \frac{(c_1^2 - 2)^3}{54\,c_1^2},\qquad
(b)\quad \D e=-\cN\,\frac{(c_1^2 - 2)^3}{27c_1^2},
\eeq
for the first and second solution respectively. Recall that
$\cN$ denotes the normalization of the standard SU(2) generators $\cN=\tr(l_i l_i)=\tr(r_i r_i)>0$ (no sum over $i$).

{\bf More solutions.} One can get more solutions for ansatzes breaking the obvious SO(3) symmetry of \refeq{lag-eff}. One can check that the following is the consistent ansatz 
\beqa\label{ans2}
X_2=X_3=t_1=0,\quad
{}[X_1,t_2]=i a\ t_3,\quad [X_1,t_3]=i b\ t_2,\quad [t_2,t_3]=i c\ X_1
\eeqa
For $c_1\geq\sqrt2$ one gets \footnote{There are also solutions which generate infinite dimensional algebras. They correspond to quantum $\RR^2$ and  shall not be discussed here.}
\beqa\label{eqm0}
(a,\ b,\ c)
=(\sfrac14(3c_1-\sqrt{16+c_1^2}),\ \sfrac14(-3c_1+\sqrt{16+c_1^2}),\
\sfrac14(3c_1+\sqrt{16+c_1^2}))
\eeqa
The formulae above define just another SU(2) algebra. 

For \refeq{eqm0} the energy above $e_0$ is
\beqa\label{en0}
\D e&=&-\sfrac\cN{32\cdot 64}
(3c_1-\sqrt{16+c_1^2}\ )^3(c_1+\sqrt{16+c_1^2}),
\eeqa
 what is negative
for $c_1>\sqrt2$. Thus we expect that \refeq{eqm0} represent local minimum.

Thus we conclude that  in all cases where the non-trivial solutions exist ($|c_1|>\sqrt{2}$) $\D e$ is negative thus total energy density is below $e_0$.  
This indicates that the solutions are stable although the analysis of this aspect of the dynamics is beyond the scope of this work.
D3-branes in the throat expand to form fuzzy $S^2$ (or $S^2\x S^2$) so the total brane world-volume is like D5-brane (or D7-brane). It is also interesting to compare \refeq{en1} with \refeq{en0}. It is easy to see that the former formula gives lower energy thus preferable configurations.


\ssec{The case of all massive moduli}\label{sec:sol-e}
Massive open string moduli are required for any phenomenologically viable string compactification. The masses can originate from different sources e.g. in case of throat geometries they may appear as due to the K\"ahler modulus stabilization and instantons \cite{KKLMMT}. One can expect that the more moduli have masses the more restrictive are conditions for the above instability to exist. 
In this part of the work we shall briefly consider the case when all the moduli have the same mass. 
Thus we add mass to $\Th_i$ modes in \refeq{lag-eff0} .

Then the ansatz \refeq{ans} does not give any non-trivial solutions, but 
\refeq{ans2} does. One finds
\beq
(a,\ b,\ c)=(x,-x,x),\qquad x=\sfrac14(c_1-\sqrt{c_1^2-16})
\eeq
The energy $\D e$  of the configuration is negative for 
$c_1< -3\sqrt{2}$.
\beq
\D e=\sfrac\cN{32\cdot 64}
(\sqrt{32}+c_1-\sqrt{c_1^2-16})(\sqrt{32}-c_1+\sqrt{c_1^2-16})
(c_1-\sqrt{c_1^2-16})^2
\eeq
Thus also in this case we get non-trivial solutions similar to those of the previous section although the required $|c_1|$ is bigger: $|c_1|> 3\sqrt{2}$.

\section{Conclusions}

In the BGMPZ throat single D3-brane breaks SUSY. 
Because $G_3$ is not ISD the 
Myers effect occurs and yields departure from a naive, single brane picture of brane stabilization. It is interesting that the phenomenon exists for all values of deviation form ISD case i.e. all values of the  $q$ parameter.
Hence
even when one brane is stabilized a multi-D3 brane system  may tend to form an extended object
(e.g. D5-branes) and lose non-abelian gauge symmetry.
 
This effect might be disastrous in models where D3-branes
provide color SU(3) degrees of freedom (see e.g. \cite{antoniadis}). The reason is that the bound states as described above break the SU(3) symmetry. Thus special care is needed for models with moduli stabilization because it might happen that single brane stabilization is not  enough to
guarantee non-abelian gauge symmetry  for multi-brane systems.

On the other hand one may speculate that a similar 
effect is responsible for
electroweak symmetry breaking. This is sounds very attractive
specially that it might occur simultaneously with  SUSY breaking as it happens for the BGMPZ background \footnote{Similarly,  D3-branes form a bound state in AISD backgrounds \cite{KPV}.}. 
Let us recall that it is generally accepted that in MSSM SUSY and electroweak symmetry breaking are tightly connected \cite{nilles,martin}\footnote{I would like to thank you Piotr Chankowski for a fruitful discussion on this point.}. If fact the former triggers the latter.

To what extend and in what cases
SUSY breaking is related to gauge symmetry breaking in string models? 
The answer to this question goes beyond the scope of this paper but let us make few remarks.
For SUSY preserving backgrounds 
$G_3$-form must be $(2,1)$  and primitive \cite{GP,gubser}. On the other hand no force condition for D3-branes imply
ISD $G_3$.  SU(3) structure manifolds allows ISD $G_3$ to have $(2,1)$ and $(0,3)$ components. Thus, in general, if one wants to relate the gauge symmetry breaking as above  with SUSY breaking the latter can not happen due to non-trivial $(0,3)$ component of $G_3$.
The bad news are that compactifications with
non-ISD $G_3$ require some special conditions to be imposed e.g. they must contain O5 or anti-O3 planes \cite{GKP}. 

Although the model considered in this work is not realistic
it is interesting that string theory can relate SUSY and gauge symmetry breaking.  
Thus it would be very interesting to build realistic string scenario along the lines presented in this work because then we could hope to immerse MSSM into a string theory connecting gauge and SUSY breaking in a nice framework.

We have not discussed here higher dimensional branes (e.g. D7-branes).
One may speculate that their behaviour could be changed by Myers effect too. If so, the extra care is needed for unification model building.\\

{\bf Acknowledgments.}
The author would like to thank Emilian Dudas and other members of the theory group in Ecole Polytechnique for discussions and hospitality. I would to thank Stefan Theisen  for discussions and for reading the manuscript. Also I would like to thank Piotr Chankowski for inspiring clarification of the MSSM and beyond models.\\

\appendix
\sec{BGMPZ geometry}
BGMPZ background is quite complicated and we are not going to describe its full form here referring to the original paper \cite{BGMPZ}.
We shell display only few formulae in order to give some hints to the reader. 
The approximate expressions for all necessary functions are given in the next subsection, while in the last subsection we shall describe an
 approximation to the geometry of the throat we are going to use.

We start with the metric
\beqa
ds^2 =
   e^{2 A } dx_{\mu} dx^{\mu} +  \sum_i^6 E_i^2  \, ,
\label{PTmetric}
\eeqa
Topologically the manifold with the metric $\sum_i^6 E_i^2$ is a certain bundle  $\RR_+\ltimes S^2\ltimes S^3$, where $\RR_+$ is radial direction parameterized by $\tau$, 
$S^2$ is parameterized by $\th_1,\phi_1$  and $S^3$ by Euler angles $\psi,\th_2,\phi_2$. 

Vielbains $E_i$ are in the form of the so-called PT ansatz \cite{PT}
\beqa \label{vielbeins}
E_1 &=& e^{\frac{x+g}{2}}  e_1 =  e^{\frac{x+g}{2}} d\th_1 \, ,\nn \\
E_2 & =& e^{\frac{x+g}{2}}  e_2 = -  e^{\frac{x+g}{2}} 
\sin\th_1 d\phi_1  \, ,\nn \\
E_3 &=& e^{\frac{x-g}{2}}  \tilde{\ep}_1 =  e^{\frac{x-g}{2}}  
(\ep_1 - a(\tau)  e_1 ) \, , \\
E_4 &=& e^{\frac{x-g}{2}}  \tilde{\ep}_2 =  e^{\frac{x-g}{2}}  
(\ep_2 - a(\tau)  e_2) \, , \nn  \\
E_5 &=& e^{-3 p - \frac{x}{2}}  dt 
\, , \nn \\ 
E_6 &=& e^{-3 p - \frac{x}{2}}  \tilde{\ep}_3 = 
e^{-3 p - \frac{x}{2}}  (\ep_3 + \cos\th_1 d\phi_1) \, , \nn
\eeqa
where $a,\ g,\ p,\ x,\ A$ are also 
functions of the radial coordinate $\tau$ only. The vielbeins are
$e_1=d\th_1$, $e_2=-\sin\th_1d\phi_1$.
Similarly $\{\ep_1,\ep_2,\ep_3\}$ are the left-invariant forms on  $S^3$
with Euler angle coordinates $\psi,\th_2,\phi_2$
\beqa \label{vielbeins2}
\ep_1 & \equiv & \sin\psi\sin\th_2 d\phi_2+\cos\psi d\th_2
\ , \nonumber  \\
\ep_2   &\equiv &  \cos\psi\sin\th_2 d\phi_2 - \sin\psi d\th_2
\ ,  \nonumber \\
\ep_3  &\equiv & d\psi + \cos\th_2 d\phi_2\ .
\eeqa
The fluxes of the PT ansatz respect the $SU(2) \times SU(2)$ symmetry. 
\beqa
H &=&  h_2  \, \tilde \ep_3 \w  (\ep_1 \w e_1 + \ep_2 \w e_2 )
+  d\tau \w \big[ h'_1  (\ep_1 \w \ep_2 + e_1 \w e_2) 
\nonumber \\ 
& & +  \chi' \,  ( e_1 \w e_2 -\ep_1 \w \ep_2)
  + h'_2 \,  (\ep_1 \w e_2 -  \ep_2 \w e_1 )\big]  \, ,\label{hhh}  \\
F_3 & = &  P \, \big[  \tilde \ep_3\w \big(  \ep_1 \w \ep_2 +  e_1 \w e_2
-  b \,   (\ep_1 \w e_2 - \ep_2 \w e_1) \big) \nonumber \\
& & +  b' \, dt \w (\ep_1 \w e_1 + \ep_2 \w e_2) \big] \, ,   \\
F_5 & =&  {\cal F}_5  +  *{\cal F}_5\ , \\\
{\cal F}_5 & = &  K \,  
e_1 \w e_2 \w \ep_1 \w \ep_2 \w \ep_3 \, . 
\label{fluxes}
\eeqa
where $P=-\frac{M}4\a'$, $K(\tau)=2P[h_1(\tau)+b(\tau)h_2(\tau)],\ 
b = -\tau(\sinh\tau)^{-1},\ h_1(\tau) = h_2(\tau) \cosh\tau $ and 
$h_2,\ \chi$ are other functions depending on $\tau$ only.

There is also non-trivial dilaton $\phi(\tau)$:
\beq
 \phi' = \frac{(C - b) (a C - 1)^2}{(b C - 1) S} e^{-2 g}
\eeq
with $C = -\cosh\tau , S = -\sinh \tau $.

\ssec{Small $q,\ \tau$ approximation}
\label{approx}
 We shall display here the near throat ($\tau\ll 1$) and near KS 
 ($q\ll 1$) approximation of functions defining the background
 (we also have $v=e^{6p+2x}$).
\beqa
&&a(\tau)=-1 + \left( \frac{1}{2} + q \right) \,\tau^2 \\
&&v(\tau)=\tau + \frac{\left( -4 + 21\,q^2 \right) \,\tau^3}{30}\\
&&e^g=\tau + \left( -\frac{1}{3}   - q - \frac{q^2}{2} \right) \,\tau^3\\
&&e^x={\tilde M}\,e^{{\phi_0}}\ \left( \frac{\tau\,\lat }{2} + 
    \frac{ -10 + 6\,{\lat }^2 + 81\,q^2\,{\lat }^2 }{180\,
       \lat }\ \tau^3 \right)\\
&&A={A_0} + 
  \frac{\tau^2}{18\,{\lat }^2}
\\
&&h_2=-\frac {\tilde M}6\ e^{{2\phi_0}}\, \tau
      \left( 1 + \left( - \frac{7}{30}   + 2\,q^2 \right) \,\tau^2 
         \right)  \\
&&\chi{}'=\frac{2\,{\tilde M}}{45}\,q\,e^{2\phi_0}\, \ \tau^2
    \left( 15 - 4\,\tau^2 + 54\,q^2\,\tau^2 \right)\\
    \label{dilaton}
&&\phi=\phi_0 + q^2 \tau^2 + \left(-\frac{q^2}6 + \frac{11 q^4}{10}\ \tau^4\right)
\eeqa
where  
$e^{\phi_0}=  {\sqrt{1 - 9\,q^2\,{\lat }^2}}$ is the value of the dilaton at the bottom of the throat in terms of deformation parameter $q$.  
${\tilde M},\ \lat,\ A_0$ are integration constants. Comparison with KS solution gives
$e^{2A_0}=\ep^{4/3}(g_s M\a'2^{1/3}a_0^{1/2})^{-1},\ 
{\tilde M}=\half g_s M\a',\ 
\lat^2= \frac{2\cdot 6^{1/3}}3\, a_0$, for $\a_0\approx 0.718$.
The expression for $a_0$ is the same as given in \cite{KS}.

\ssec{Geometry of spheres in BGMPZ}
\label{sec:flatS}
In this subsection we shall explain in what sense the 
geometry of the bottom of the throat is $\RR^6$.

The throat contains the bundle $S^2\ltimes S^3$. 
The latter was nicely described in \cite{MT}.
One defines
\beq 
L_i=\twomat{\cos(\frac{\th_i}2)\
e^{i\,\left( {\psi}/{2} + \phi_i\right)/2}
}
{-\sin(\frac{\th_i}2)\
e^{{-i}\,\left({\psi}/{2} - \phi_i\right)/2}
} 
{\sin(\frac{\th_i}2) \
e^{{i}\,\left({\psi}/{2} -\phi_i\right)/2}
}
{\cos(\frac{\th_i}2)\
e^{{-i}\,\left({\psi}/{2} + \phi_i\right)/2}
}=e^{i\si^3\phi_i/2}e^{-i\si^2\th_i/2}e^{i\si^3\psi/4}
\eeq
and 
\beq
T=L_1\si^1L_2^+\si^1
\eeq
The matrix $T\in SU(2)$, thus it is $S^3$. We represent it as follows
\beq
T=e^{i\si^2 \pi/4}\
e^{i\si^3\phi/2}e^{-i\si^2(\pi/4+\th/2)}e^{i\si^3\chi/2}\in S^3
\eeq
In this paper we parameterize $S^2\ltimes S^3$ with $\th_2,\phi_2$ and $\phi,\th,\chi$ defined as above.

Now let us discuss 6d metric close to $\tau=0$. It has the form:
\beq
ds_6^2=d\tau^2+d\Om_3^2+\frac{\tau^2}2(g_1^2+g_2^2)
\eeq
where $d\Om_3^2$ is the metric on $S^3$ represented by $T$. Close to $\th=\chi=\phi=0$ the metric is flat
\beq\label{s3}
d\Om_3^2=d\th^2+d\phi^2+d\chi^2\equiv (d\Th_i)^2
\eeq
The metric on $S^2$ fibered over $S^3$ is.
\beqa
&&(g_1^2+g_2^2)=\\
&&~~~~\half[\left( \sin\psi \,d\th_2 - 
        \sin {{\th}_1}\,d\phi_1 - 
        \cos \psi\,\sin \th_2\,d\phi_2\right)^2
	+ \left( d\th_1- 
        \cos \psi\,d\th_2 - 
        \sin \psi\,\sin \th_2\,d\phi_2\right)^2]\nn
\eeqa
Because $\psi$ is small and $\phi_1\approx -\phi_2, \ \th_1\approx \th_2$ then we get
\beq
(g_1^2+g_2^2)=2(\sin^2\th_2\,d\phi_2^2+d\th_2^2)
\eeq
In this approximation the space parameterized by $(\tau, \th_2,\phi_2)$ is $\RR^3$.
All together the 6d metric  is
\beq
ds_6^2=(d\Th_i)^2+(dX_i)^2, \quad i=1,2,3
\eeq
Here we have denoted the local coordinates on $\RR^3$ by
$X_i$. Notice that all of these coordinates are dimensionless.

We also need formulae for $B$ and $\om_{FH}$ 2-forms appearing in 
\refeq{myers-my}. Both are proportional to the following form
\beqa
\tau(e_1+e_3)\w(e_2+e_4)
=\sfrac12 \ep_{ijk}X_i\ d\Th_j\w d\Th_k
\eeqa

\newpage

\end{document}